\documentclass[journal]{IEEEtran}
\usepackage{cite}

\ifCLASSINFOpdf
  \usepackage[pdftex]{graphicx}
  \DeclareGraphicsExtensions{.pdf,.jpeg,.png}
\else
\fi
\usepackage{bm} 
\usepackage{amssymb}
\usepackage{amsmath}

\begin{document}
%
\title{Design principles of apodized grating couplers}
%
%
%

\author{Zhexin~Zhao,~
        and~Shanhui~Fan,~\IEEEmembership{Fellow,~OSA,~Fellow,~IEEE}
\thanks{Z. Zhao and S. Fan are with the Ginzton Laboratory, 348 Via Pueblo, Stanford University, Stanford, CA 94305, USA. e-mail: shanhui@stanford.edu.}}
\maketitle

\begin{abstract}
To optimize the coupling efficiency of grating couplers, we extend an analytical model for the design of apodized grating couplers, taking into account the constraints on the upper and lower bounds of the scattering strength as determined by fabrication technology. We prove that our model gives the global optimum distribution of the scattering strength with the presence of such constraints. We apply our model to the standard problem of coupling from a silicon chip to a single-mode fiber, as well as more complex problems including the coupling to a vortex beam, and the design of a focusing grating. In the problem of coupling to a vortex beam, we validate our model with full-wave numerical simulations. For this problem, our design obtains efficiency that is significantly higher compared with existing designs. Our theoretical study provides a solid foundation for apodized grating coupler design in different material platforms, and our deterministic algorithm can function as a near-optimum starting point for advanced optimizations.
\end{abstract}


%
\IEEEpeerreviewmaketitle

%
%
%
%

\section{Introduction}
\label{secintro}
\IEEEPARstart{T}{he} grating coupler is a crucial component in integrated photonics which couples light into or out from the integrated photonic circuits \cite{mekis2010grating, marchetti2019coupling}. Its coupling efficiency is important to the performance of the whole system. Considerable efforts have been devoted into designing grating couplers with high coupling efficiency \cite{taillaert2004compact, chen2010apodized, marchetti2017high, wang2005embeddedslant, roelkens2006highoverlay, su2018fully, verslegers2019method, ding2014fully, bozzola2015optimising, michaels2018inverse}. A uniform grating coupler has an exponential scattering intensity distribution and, consequently, a maximum theoretical coupling effieicny of 80\% with a Gaussian beam, due to the mismatch between an exponential distribution and a Gaussian distribution \cite{orobtchouk2000high}. 
Therefore, to achieve higher coupling efficiency, the grating couplers must be apodized, which means that the geometric parameters like periodicity and etch length are changed along the grating coupler. 

A well-established description of the apodized grating coupler is to model the grating as a continuous scatterer with a position-dependent scattering strength \cite{mekis2010grating, taillaert2004compact}. With this model, 
the ideal scattering strength can be expressed as a function of the target output. Using a mapping between the scattering strength and the geometric parameters of the grating, a concrete design of the apodized grating coupler can be found.
Guided by this model, grating couplers with high coupling efficiency and near-unity mode-matching efficiency have been demonstrated \cite{taillaert2004compact, chen2010apodized, ding2014fully}.  

However, the previous model, though successful, has an important limitation. The optimized scattering strength predicted by the previous model, which is referred to as the {ideal scattering strength}, is a continuous function starting from zero and usually unbounded \cite{mekis2010grating}. But in practice, the scattering strength can neither be arbitrarily large due to the finite scattering strength of the grating teeth, nor be arbitrarily small, due to the minimal feature size allowed in fabrication. 

In this study, we extend the previous model by including the constraints of upper and lower bounds of the scattering strength. We provide a formalism that can determine the globally optimal scattering strength under these constraints. 
This extension results in modest improvement in coupling efficiency for the ``standard'' grating couplers in silicon photonics coupling with single mode fiber. However, our extension becomes important for grating couplers coupling with more complex beams \cite{demirtzioglou2019apodized, zhou2019ultra, liu2016chip, nadovich2016forked, nadovich2017focused}, such as optical beams with non-zero orbital angular momentum, or in integrated photonic systems using materials other than silicon \cite{krasnokutska2019high, chen2016high, maire2008high}.   

The paper is organized as following: A brief review of the coupling efficiency and the previous {ideal model} is presented in Sec. \ref{sec:review}. In Sec. \ref{sec:theory}, we discuss our extension of the {ideal model} and provide an algorithm to obtain the scattering strength. As a validation of our model, in Sec. \ref{sec:design_procedure}, we present our design procedure with a illustrative design of a grating coupler on a silicon-on-insulator (SOI) platform, which has coupling efficiency comparable to the state-of-the-art. We extend both the {ideal model} and our extension to two-dimensional non-focusing gratings in Sec. \ref{sec:2d_nonfocus}. In Sec. \ref{sec:design_oam}, we analytically design and numerically demonstrate a highly efficient grating coupler coupling to a Laguerre-Gaussian beam carrying orbital angular momentum. We extend our model to the design of a fan-shape focusing grating coupler in Sec. \ref{sec:2d_focusing}. In Sec. \ref{sec:implication} we further discuss some implications of our model, such as how the upper and lower bounds of the scattering strength constrain the upper limit of the coupling efficiency. We conclude in Sec. \ref{sec:conclusion}.

\section{Review of the {ideal model}}
\label{sec:review}
In this section, we outline the description of the coupling efficiency of the grating coupler system, and briefly review the previous {ideal model} that provides the {ideal scattering strength}. 

\begin{figure}
    \centering
    \includegraphics[width = 0.76\linewidth]{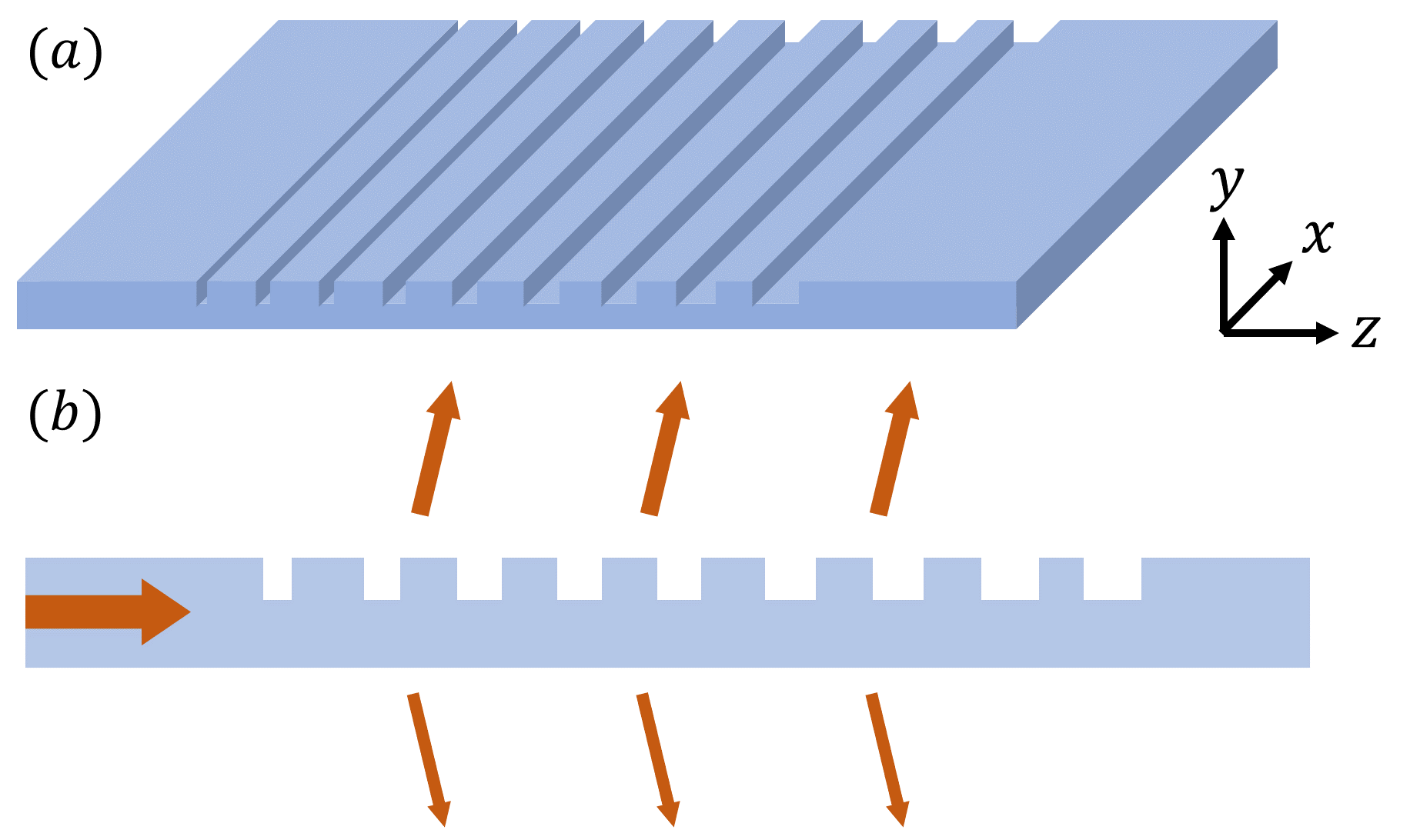}
    \caption{Schematic of an apodized grating coupler (a) and its cross section (b).}
    \label{fig:schematic}
\end{figure}

Consider a grating coupler that couples the power in the guided mode of a waveguide to a target mode profile (Fig. \ref{fig:schematic}). The coupling efficiency ($\eta$) can be obtained through an overlap integral \cite{michaels2018inverse}:
\begin{equation}
\label{eq:couplingefficiency}
    \eta = \frac{1}{P_{\textrm{wg}}P_\textrm{t}}\Big|\iint \frac{1}{2} \bm{E}\times \bm{H}_\textrm{t}^* \cdot d\bm{S} \Big|^2,
\end{equation}
where $P_\textrm{wg}$ and $P_\textrm{t}$ are the power in the guided wave and in the target mode respectively, $\bm{E}$ is the electric field scattered by the grating coupler, and $\bm{H}_\textrm{t}$ is the magnetic field of the target mode profile. The overlap integral can be carried out above the grating coupler. Equation (\ref{eq:couplingefficiency}) suggests that to maximize the coupling efficiency, the polarization, amplitude distribution and phase distribution of the scattering field should match that of the target mode. 
In the setup of Fig. \ref{fig:schematic}, which is a two-dimensional system having translational symmetry along the $x$-direction, the polarization matching is guaranteed by symmetry.
The phase matching is also well-studied. For a uniform grating, it is achieved by choosing the grating pitch to compensate the momentum mismatch between the guided mode and the target output beam \cite{mekis2010grating, marchetti2019coupling}. For an apodized grating, the phase matching can be satisfied with small adjustments of the grating pitch \cite{michaels2018inverse}.
Assuming that the polarization matching and phase matching can be achieved, we focus on the matching between the amplitude distributions. Under this assumption, the coupling efficiency becomes
\begin{equation}
    \label{eq:couplingefficiency2}
    \eta = \frac{1}{P_{\textrm{wg}}P_\textrm{t}}\Big|\int dx \int dz A(x, z) A_\textrm{t}(x, z) \Big|^2,
\end{equation}
where $|A(\bm{r})|^2$ represents the intensity of the field at $\bm{r}$. The guided wave is propagating along $z$-direction, the $y$-direction is perpendicular to the grating coupler, and the remaining transverse direction is the $x$-direction (Fig. \ref{fig:schematic}). 

When the grating teeth are parallel, as in most common cases, the $x$- and $z$-dependence of the amplitude are separable, i.e. $A(x, z) = A_\textrm{x}(x)A_\textrm{z}(z)$. (We can choose $A_\textrm{x}(x)$ such that it is dimensionless and takes peak value of unity.) Furthermore, for target modes like Gaussian beams, the decomposition $A_\textrm{t}(x, z) = A_\textrm{tx}(x)A_\textrm{tz}(z)$ also holds. Since the amplitude matching in the transverse direction is straightforward \cite{taillaert2004compact}, the remaining challenge is to design the apodized grating coupler to maximize the following overlap integral along the longitudinal $z$-direction:
\begin{equation}
    \label{eq:overlapz}
    \eta_\textrm{z} = \frac{1}{P_\textrm{wg} P_\textrm{t}}\Big|\int dz A_\textrm{z}(z) A_\textrm{tz}(z) \Big|^2.
\end{equation}

Equation (\ref{eq:overlapz}) is equivalent to the coupling efficiency of a one-dimensional grating coupler system, where both the grating coupler and the target mode extend uniformly along $x$-direction. The power terms like $P_\textrm{wg}$ and $P_\textrm{t}$ take the unit of power per unit length along the $x$-direction.
Due to Cauchy-Schwarz inequality
\begin{align}
    \label{eq:CauchySchwarz}
    \eta_\textrm{z} \leq \frac{\int dz |A_\textrm{z}(z)|^2 \int dz |A_\textrm{tz}(z)|^2}{P_\textrm{wg} P_\textrm{t}},
\end{align}
unity coupling efficiency is achieved only if the amplitude matching is achieved and the guided power is entirely scattered to one side of the grating coupler, i.e.
\begin{align}
    \label{eq:CauchySchwarz_A}
    A_\textrm{z}(z) & = C_1 A_\textrm{tz}(z) ,\\
    \label{eq:CauchySchwarz_Pwg}
    P_\textrm{wg} & = \int dz |A_\textrm{z}(z)|^2,
\end{align}
where $C_1$ is a constant, and $P_\textrm{t} = \int dz |A_\textrm{tz}(z)|^2$ is always satisfied. 

Based on the discussion above, the well-established model, which we refer to as the \textit{ideal model} in this paper, for such one-dimensional grating coupler system is summarized as following. The apodized grating coupler (Fig. \ref{fig:schematic}) is modeled as a continuous scatterer whose scattering strength depends on $z$, such that
\begin{align}
    \label{eq:alpha_power_decay}
    \frac{dP(z)}{dz} = - 2\alpha(z) P(z),
\end{align}
where $P(z)$ is the remaining power in the guided mode and $\alpha(z)$ is the scattering strength of the grating. The {ideal model} also assumes no reflection in the grating coupler region. The intensity of the scattering light is 
\begin{equation}
    \label{eq:scattered_light_1}
    S(z) = 2\alpha(z)P(z).
\end{equation}
To maximize the coupling efficiency, the amplitude matching condition (\ref{eq:CauchySchwarz_A}) should be satisfied, such that $S(z) = S_\textrm{t}(z)$, where $S_\textrm{t}(z)$ is the target intensity distribution. Together with (\ref{eq:alpha_power_decay}) and (\ref{eq:scattered_light_1}), we find that
\begin{align}
    \frac{dP(z)}{dz} & = -S_\textrm{t}(z), \\
    \label{eq:P_1} P(z) & = P(-\infty) - \int_{-\infty}^z S_\textrm{t}(t) dt, \\
    \label{eq:alpha_1} \alpha(z) & = \frac{1}{2}\frac{S_\textrm{t}(z)}{P(-\infty) - \int_{-\infty}^z S_\textrm{t}(t) dt},
\end{align}
where $P(-\infty) = P_\textrm{wg}$ and $-\infty$ ($+\infty$) denotes any position before (after) the grating coupler region. Equation (\ref{eq:alpha_1}) resembles Eq. (3) in \cite{taillaert2004compact}.

Furthermore, if the guided power is entirely extracted, $P(-\infty) = \int_{-\infty}^{+\infty} S_\textrm{t}(t)dt$. Therefore, the scattering strength is
\begin{equation}
    \label{eq:alpha_complete_scattering}
    \alpha(z) = \frac{1}{2}\frac{S_\textrm{t}(z)}{\int_{z}^\infty S_\textrm{t}(t) dt},
\end{equation}
which is consistent with Eq. (3) in \cite{mekis2010grating}. 

On the other hand, if only a portion ($\zeta$) of the guided power is scattered by the grating coupler, 
\begin{equation}
    \zeta P(-\infty) = \int_{-\infty}^{+\infty} S_\textrm{t}(t)dt, 
\end{equation}
the scattering strength (\ref{eq:alpha_1}) becomes
\begin{equation}
\label{eq:alpha_partial_scattering}
    \alpha(z) = \frac{1}{2}\frac{\zeta S_\textrm{t}(z)}{\int_{-\infty}^\infty S_\textrm{t}(t)dt - \zeta \int_{-\infty}^z S_\textrm{t}(t) dt},
\end{equation}
which is used in \cite{mehta2017precise, nadovich2017focused}. This scattering strength is regarded as the optimum for a fixed scattered portion $\zeta$, but obviously not the optimal scattering strength in general, since part of the power remains in the waveguide (i.e. $\zeta \neq 1$). In this paper, we refer to the scattering strength defined in either (\ref{eq:alpha_complete_scattering}) or (\ref{eq:alpha_partial_scattering}) as the \textit{ideal scattering strength}.

From (\ref{eq:alpha_complete_scattering}) and (\ref{eq:alpha_partial_scattering}), we find that multiplying $S_\textrm{t}(z)$ by a constant factor has no influence on the optimal scattering strength. Thus, we can simply take $S_\textrm{t}(z) = |A_\textrm{tz}(z)|^2$. 
Moreover, if only a portion ($\xi$) of the guided power is scattered to the target mode side, where $\xi$ is independent of $z$, the optimal scattering strength is unchanged, and the optimal coupling efficiency is reduced by a factor of $\xi$.
We also find that the numerator and denominator in (\ref{eq:alpha_complete_scattering}) both approach $0$ as $z$ increases. When the target mode is a Gaussian beam, $\alpha(z)$ is a monotonic increasing function of $z$ \cite{mekis2010grating}. However, the scattering strength cannot be arbitrarily large and (\ref{eq:alpha_partial_scattering}) is sometimes used to ensure that $\alpha(z)$ is upper bounded \cite{nadovich2017focused}, because $\alpha(z)$ given by (\ref{eq:alpha_partial_scattering}) approaches 0 as $z$ approaches infinity for $\zeta < 1$, and, therefore, is upper bounded. 
Due to this property, (\ref{eq:alpha_partial_scattering}) is used when the fidelity of the output mode, which is the similarity between the output mode and the target mode, is more important than the coupling efficiency \cite{mehta2017precise, nadovich2017focused}.

\section{Theory and Algorithm}
\label{sec:theory}

The previous {ideal model} reviewed in Sec. \ref{sec:review} generally gives an {ideal scattering strength} that starts from zero and increases to a large value for a long grating. 
In this section, we present our extension to the previous {ideal model} by taking the upper and lower bound of the scattering strength into account. 

We consider a grating coupler with length $L$ whose scattering strength is upper- and lower-bounded ($\alpha \in [\alpha_{\textrm{min}}, \alpha_{\textrm{max}}] \cup \{0\}$).
Below we refer to such a range of $\alpha$ as the \textit{feasible range}. The scattering strength certainly cannot be arbitrarily large for a given index contrast in a grating. Also, since there is usually a constraint on the minimum feature size of the grating, the scattering strength cannot be arbitrarily small either if it is non-zero. On the other hand, it can take the value of zero, which corresponds to a uniform region with no grating. Here, by ``lower bound'', we refer only to non-zero values of $\alpha$. 
As a starting point, in this section, we consider only the one-dimensional grating coupler system and focus only on the amplitude matching.
With these simplifications, the coupling efficiency is 
\begin{equation}
    \label{eq:coupling_finite_length}
    \eta = \xi \Big|\int_0^L \sqrt{S(z)} A_\textrm{t}(z)dz \Big|^2,
\end{equation}
where $\xi$, describing the directivity of the grating \cite{marchetti2019coupling}, is the portion of light scattered towards the side of the target mode, and the grating coupler is placed in $z \in [0, L]$. Comparing with (\ref{eq:overlapz}), we set the power in the guided mode at $z=0$ to be unity. The amplitude of the target mode is normalized such that $P_\textrm{t}=1$, and the subscript $z$ is omitted for simplicity.

The optimized grating coupler, characterized by $\alpha(z)$, should maximize the coupling efficiency. In practice, the portion of light that is scattered towards the target side of the grating is determined by the technology, such as layer thicknesses and etching depth \cite{wang2005embeddedslant, bozzola2015optimising, marchetti2017high, roelkens2006highoverlay}, and can be optimized separately. Thus, we take $\xi$ as a constant independent of the grating apodization.
With a $z$-dependent scattering strength, the power remaining in the guided mode and the scattering intensity are obtained from (\ref{eq:alpha_power_decay}) and (\ref{eq:scattered_light_1}):
\begin{align}
    \label{eq:power_remaining}
    P(z) & = \exp\Big[ -2\int_0^z \alpha(t)dt \Big], \\
    \label{eq:scattered_light_2}
    S(z) & = 2\alpha(z)\exp\Big[ -2\int_0^z \alpha(t)dt \Big].
\end{align}
Since both $S(z)$ and $A_\textrm{t}(z)$ are real and $\xi$ is a constant, we find that to maximize the coupling efficiency is equivalent to maximize the term in the absolute sign of (\ref{eq:coupling_finite_length}):
\begin{equation}
    \label{eq:f}
    f = \int_0^L \sqrt{2\alpha(z)} \exp\Big[ -\int_0^z \alpha(t)dt \Big] A_\textrm{t}(z) dz.
\end{equation}
Since $f$ is a functional of $\alpha(z)$, the optimization problem can therefore be formulated as:
\begin{align}
    \label{eq:optimization}
    \begin{split}
        \textrm{Objective:} \; & \textrm{max} \, f(\alpha(z)), \\
        \textrm{Subject to:} \; & \alpha(z) \in [\alpha_{\textrm{min}}, \alpha_{\textrm{max}}]\cup\{0\}.
    \end{split}
\end{align}

We now prove that the constraint optimization problem has a global optimum given a target mode, by first showing that the problem exhibits optimal substructure \cite{cormen2009introduction}. 
Optimal substructure refers to the fact that the optimal scattering strength for the whole grating must also be the optimal scattering strength for any tail potion of the grating coupler, i.e. the grating segment $[z, L]$ for any $z$. The existence of the optimal substructure allows us to derive a recursive relation to determine the optimal scattering strength, since the optimal scattering strength at position $z$ depends only on the target mode and the optimal scattering strength after $z$. Therefore, we can determine the optimal scattering strength from the end to the beginning of the grating, and we show that the derived scattering strength is the global optimum.

To illustrate the optimal substructure of the optimization problem, 
we separate $f$ into two parts:
\begin{align}
\label{eq:f_separation}
    f & = f^-(z) + \exp \Big [{-\int_0^{z}\alpha(t)dt} \Big ] f^+(z),
\end{align}
where
\begin{align}
    \label{eq:f_minus}
    f^-(z) & = \int_0^{z} \sqrt{2\alpha(s)} \exp\Big[{-\int_0^s \alpha(t)dt}\Big] A_\textrm{t}(s) ds, \\
    \label{eq:f_plus}
    f^+(z) & = \int_{z}^L \sqrt{2\alpha(s)} \exp \Big[{-\int_z^s \alpha(t)dt}\Big] A_\textrm{t}(s) ds.
\end{align}
$f^-(z)$ depends only on the grating scattering strength between 0 and $z$, while $f^+(z)$ depends only on the grating scattering strength between $z$ and $L$. Comparing (\ref{eq:f}) and (\ref{eq:f_plus}), we find that $f = f^+(0)$. Also, $\xi|f^+(z)|^2$ has the physical meaning of the coupling efficiency of the grating coupler with grating region restricted to $[z, L]$. To maximize $f$, the scattering strength must also maximize $f^+(z)$ for any $z \in [0, L]$. This can be straightforwardly proved by counter-evidence (Appendix \ref{app:optimal_substructure}). Thus, the optimization problem has optimal substructure.  

Due to the optimal substructure, we can adopt a recursive approach to study the sub-problem of finding the optimal scattering strength in region $[z, L]$ to maximize $f^+(z)$. 
The relevant question is: Assuming the optimal scattering strength in $[z+\Delta z, L]$ has been found to be $\alpha^\star(s)$ for all $s \in [z+\Delta z, L]$ such that $f^+(z+\Delta z)$ is maximized, what is the optimal $\alpha(s)$ in $s \in [z, z+\Delta z]$ such that $f^+(z)$ is maximized? Since $\Delta z$ is infinitesimal, we can approximate the scattering strength in $[z, z+\Delta z]$ as a constant $\alpha_z$. Therefore, $f^+(z)$ is a function of $\alpha_z$ only. And hence $\alpha_z$ can be determined. Thus, we get the optimal scattering strength in region $[z, L]$. In fact, one can further show that this function is concave, and reaches its maximum either at the extreme point, or at the boundaries of the feasible range.
With this recursive approach starting from the end of the grating, we can get the optimal scattering strength for the whole grating coupler. The above derivation also proves that the obtained scattering strength is a global optimum.

We proceed to give the explicit recursive relation to obtain the optimal scattering strength in $[z, z+\Delta z]$.
The dependence of $f^+(z)$ on $\alpha_z$ is:
\begin{align}
\label{eq:f_plus_deltaz}
    \begin{split}
        f^+(z; \alpha_z) = \,& \sqrt{2\alpha_z} \int_0^{\Delta z} \exp(-\alpha_z t) A_\textrm{t}(z + t) dt \\ \,& + \exp(-\alpha_z \Delta z)f^+(z + \Delta z)
    \end{split}
\end{align}
Taking derivative with respect to $\alpha_z$, where we assume $\alpha_z >0$ and postpone treating the case $\alpha_z=0$, we find
\begin{align}
    \label{eq:f_plus_derive_1}
    \begin{split}
        \frac{\partial f^+(z; \alpha_z)}{\partial \alpha_z} = &\, \Big[ \frac{1}{\sqrt{2 \alpha_z}} A_\textrm{t}(z) - f^+(z + \Delta z)\Big]\Delta z \\ & \, + O[(\Delta z)^2] ,
    \end{split}\\
    \label{eq:f_plus_derive_2}
    \frac{\partial^2f^+(z; \alpha_z)}{\partial \alpha_z^2} = &\, -(2\alpha_z)^{-\frac{3}{2}} A_\textrm{t}(z) \Delta z + O[(\Delta z)^2].
\end{align}
As $\Delta z \to 0$, $O[(\Delta z)^2]$ terms are negligible comparing with the terms proportional to $\Delta z$. To maximize $f^+(z)$, we set $\partial f^+(z) /\partial \alpha_z = 0$ and find the condition for the extreme point:
\begin{equation}
\label{eq:alphaz_extreme}
    \frac{1}{\sqrt{2 \alpha_z}} A_\textrm{t}(z) = f^+(z).
\end{equation}
The second derivative (\ref{eq:f_plus_derive_2}) is negative, which implies that $f^+(z)$ is a concave function of $\alpha_z$ and the extreme point is a maximum.
Thus, without considering the feasible range, the optimal scattering strength is:
\begin{align}
    \label{eq:alpha_opt}
    \alpha_z = \frac{A_\textrm{t}^2(z)}{2 \Big\{\int_{z}^L \sqrt{2 \alpha^\star(s)}\exp \Big[ -\int_{z}^s \alpha^\star(t)dt \Big] A_\textrm{t}(s) ds \Big\}^2},
\end{align}
where the superscript $\star$ indicates the optimum within the feasible range. Since $\alpha(z)$ should not have singularities, the integrations starting from $z$ in (\ref{eq:alpha_opt}) are equivalent to those starting from $z^+$. Equation (\ref{eq:alpha_opt}) suggests that $\alpha_z$ depends on $\alpha^\star(s)$ for $s \in (z, L]$, but not $s \in [0, z)$. Hence, (\ref{eq:alpha_opt}) can be regarded as a recursive relation. Furthermore, since $f^+(z)$ is a concave function of $\alpha_z$, the maximum is at the boundary if $\alpha_z \notin [\alpha_\textrm{min}, \alpha_\textrm{max}]$. 
Moreover, as $\alpha(z)=0$ is also eligible, one need to compare $f^+(z; \alpha_z=\alpha_\textrm{min})$ and $f^+(z; \alpha_z=0)$ to choose between $0$ and $\alpha_\textrm{min}$ if $\alpha_z$ obtained from (\ref{eq:alpha_opt}) is smaller than $\alpha_\textrm{min}$. In summary, the optimal scattering strength at $z$ within the feasible range is
\begin{align}
    \label{eq:alphaconstraint}
    \alpha^\star(z) = 
    \begin{cases} 
      0 & \alpha_z <  \alpha_{\textrm{min}} \textrm{, } f^+(z; \alpha_\textrm{min}) \leq f^+(z; 0) \\
      \alpha_\textrm{min} & \alpha_z <  \alpha_{\textrm{min}} \textrm{, } f^+(z; \alpha_\textrm{min}) > f^+(z; 0) \\
      \alpha_z &  \alpha_{\textrm{min}} \leq \alpha_z \leq  \alpha_{\textrm{max}} \\
       \alpha_{\textrm{max}} &  \alpha_{\textrm{max}} < \alpha_z
   \end{cases}.
\end{align}

Equations (\ref{eq:alpha_opt}) and (\ref{eq:alphaconstraint}) are the recursive relation for the optimal scattering strength. 
In the numerical implementation of solving $\alpha(z)$, we take finite but small step $\Delta z$.
We discretize the interval $[0, L]$ into N equal segments, and denote $\Delta z = L/N$, $z_i = \frac{i}{N}L$, $\alpha_i = \alpha_{z_i}$, $\alpha_i^\star = \alpha^\star(z_i)$, and $A_i = A(z_i)$, where $i = 0, 1, \hdots, N$. The discretized version of the extreme point condition (\ref{eq:alphaz_extreme}) is 
\begin{align}
    \label{eq:alpha_extreme_discrete}
    \begin{split}
    & \, \frac{1}{\sqrt{2\alpha_i}}A_i =  \frac{1}{2}\sqrt{2\alpha_i}A_i\Delta z \\
    & + \sum_{j=i+1}^{N-1}\sqrt{2\alpha_j^\star}\exp  \big[-\big(\frac{\alpha_i}{2} + \sum_{k=i+1}^{j-1}\alpha_k^\star + \frac{\alpha_j^\star }{2}\big) \Delta z \big] A_j \Delta z \\
    & + \frac{1}{2}\sqrt{2\alpha_N^\star} \exp \big[-\big( \frac{\alpha_i}{2} + \sum_{k=i+1}^{N-1} \alpha_k^\star + \frac{\alpha_N^\star}{2} \big) \Delta z \big] A_N \Delta z.
    \end{split}
\end{align}
for $i = 0, 1, \hdots, N-1$. 
Equation (\ref{eq:alpha_extreme_discrete}) is a transcendental equation of $\alpha_i$, which can be solved iteratively usually with only a few iteration steps. 
The grating strength at the end of the grating coupler must be as large as possible in order to minimize the remaining guided power.
So,
\begin{equation}
    \label{eq:alpha_N}
    \alpha^*_N = \alpha_{\textrm{max}}.
\end{equation}
Thus, we can numerically solve (\ref{eq:alpha_extreme_discrete}) from $i = N-1$ to $i = 0$. 
For each $i$, after solving (\ref{eq:alpha_extreme_discrete}), we obtain the optimum $\alpha_i^\star$ based on the condition (\ref{eq:alphaconstraint}), such that $\alpha_i^\star \in [\alpha_\textrm{min}, \alpha_\textrm{max}]$ or $\alpha_i^\star = 0$.

As a final remark, the position of the target beam is usually also a design parameter, which can be taken into account in our model straightforwardly. Since the maximal $f$ can be obtained for any target beam position ($z_0$), searching for the optimal $z_0$ is a single variable optimization problem. It can be solved by either brute force or gradient decent. 

In summary, we extend the {ideal model} by considering explicitly the upper and lower bounds of the scattering strength. We prove that the scattering strength given by (\ref{eq:alpha_opt}) and (\ref{eq:alphaconstraint}) is the optimum. An algorithm is also provided for solving $\alpha(z)$ numerically.

\section{Design procedure}
\label{sec:design_procedure}

In this section, we outline the deterministic design procedure towards generating a highly efficient grating coupler corresponding to available fabrication technology. This procedure is similar to previous work with the {ideal model} \cite{taillaert2004compact, marchetti2019coupling}. We repeat it for completeness and discuss details of maintaining the phase matching.
For illustration, we demonstrate the design procedure by designing a one-dimensional grating coupler (Fig. \ref{fig:schematic}(b)).

\textit{Step 1: Specify technology and experimental conditions.} 

The fabrication technology and experimental conditions constrain the achievable parameters for the design. Such parameters include waveguide core material, core thickness, top and bottom cladding material and thickness, and minimal feature size. The industry-standard testing platform gives further requirements for the fiber incident angle. 
For demonstration, we choose a silicon-on-insulator platform with 220 nm silicon thickness and 2 $\mu$m bottom oxide thickness. The top oxide cladding is assumed to be thick. The etch depth is 70 nm and the minimal feature size is 80 nm, which can be achieved by electron beam lithography or deep ultra-violet photo lithography \cite{selvaraja2009fabrication, hu2004sub}. The single mode fiber incident angle is 10$^\circ$ in air, which corresponds to $\theta=6.9^\circ$ in silicon dioxide cladding. The free space wavelength is 1550 nm. The target mode in the single-mode fiber can be approximated as a Gaussian beam with beam waist $w_0 = 5.2$ $\mu$m and we consider TE polarization (electric field polarized in $x$-direction). 
This fabrication technology and experimental condition are chosen since they are close to the industrial standard. Nevertheless, our model and design procedure can be applied to other technologies. The parameters of the fabrication technology can also be optimized by comparing the optimal grating couplers under different technologies.

\begin{figure}
    \centering
    \includegraphics[width = 0.86\linewidth]{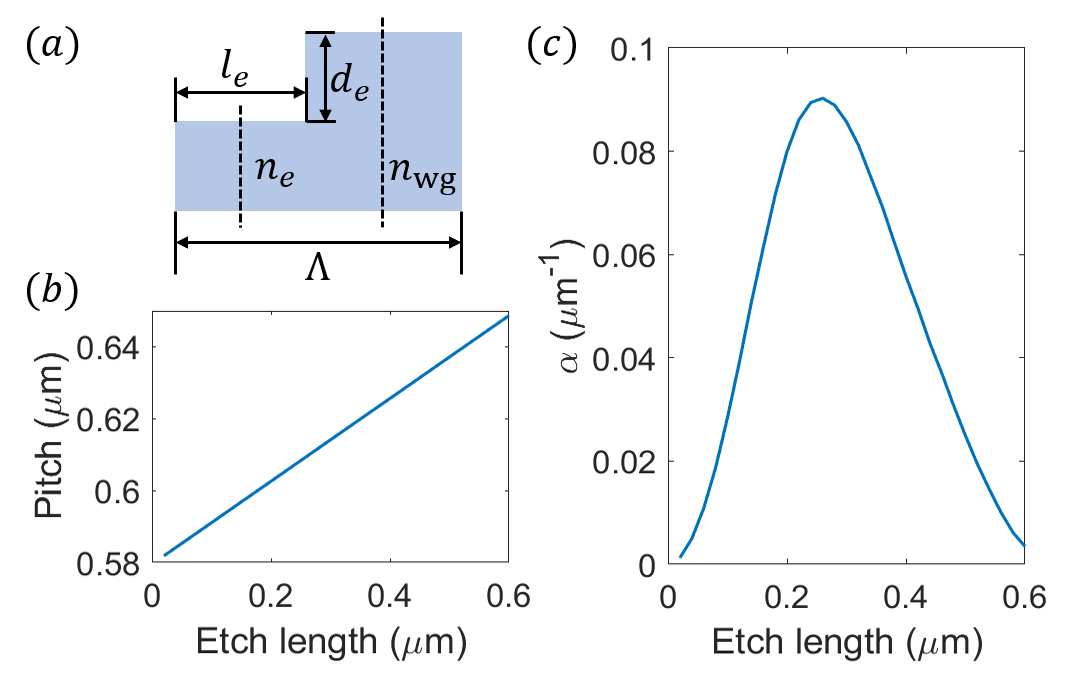}
    \caption{(a) Schematic of a unit cell in a grating coupler.  $\Lambda$ is the pitch of the grating, $l_e$ is the etch length, $d_e$ is the etch depth. The effective indices of the guided modes in the etched part and unetched part are labeled as $n_{e}$ and $n_\textrm{wg}$. (b) Pitch as a function of the etch length that satisfies the phase matching condition (\ref{eq:pitch_analytical}). (c) Grating scattering strength as a function of the etch length. (b) and (c) are for 220 nm thick silicon slab, with 70 nm etch depth, surrounding by silica.}
    \label{fig:mapping}
\end{figure}

\textit{Step 2: Generate mapping between scattering strength and grating structures.} 

A mapping from the geometrical parameters of a single unit cell of the grating to the scattering strength $\alpha$ is required for the design.
Although sophisticated scattering elements can be used in a single unit cell \cite{ding2014fully, chen2016high, halir2009waveguide, purwaha2019broadband, qin2012high, watanabe2017perpendicular}, 
as a simple demonstration, we choose the scattering element per unit cell as a single etch trench (Fig. \ref{fig:mapping}(a)). The relevant parameters are then the etch length, and the length of the unit cell which corresponds to the pitch of the grating.

For a given etch length, the length of the unit cell is determined by the phase-matching condition as follows. 
Assume that the accumulated phase per unit cell for the guided wave is $ \phi_{\Lambda} =  2\pi n_{e}l_e/{\lambda} + 2\pi n_\textrm{wg}(\Lambda - l_e)/{\lambda}$,
where $\Lambda$ is the pitch, $l_e$ is the etch length, $\lambda$ is the free space wavelength, $n_{e}$ and $n_\textrm{wg}$ are the effective indices of the etched and unetched slab waveguide \cite{michaels2018inverse}. To satisfy the phase matching condition, $\phi_\Lambda = 2\pi + 2\pi n_c \Lambda \sin\theta / \lambda$, where $n_c$ is the refractive index of the cladding, and $\theta$ is the target beam propagation angle in the cladding layer. With these assumptions, the grating pitch is 
\begin{equation}
    \label{eq:pitch_analytical}
    \Lambda = \frac{\lambda + l_e (n_\textrm{wg} - n_{e})}{n_\textrm{wg} - n_c \sin{\theta}},
\end{equation}
which is illustrated in Fig. \ref{fig:mapping}(b) for the chosen fabrication technology. This prediction of grating pitch works well empirically, especially for shallow etch or low index contrast between core and cladding.

We then numerically simulate a long uniform grating for each etch length and the corresponding pitch.
The scattering strength is extracted by fitting the remaining power in the guided mode along the propagation direction to an exponential decay.
In such a simulation, the grating should be long enough such that the guided power at the end of the grating is negligible. The mapping between the etch length and scattering strength is shown in Fig. \ref{fig:mapping}(c).

The simulation of the uniform grating also provides information of the emission phase of the grating with a specific etch length. The emission phase describes the phase of the scattered field, which can be extracted from the following field overlap integral with the target mode:
\begin{equation}
    \label{eq:emission_phase}
    f_{\textrm{e}} = \frac{1}{2} \int E(z)\times H_{\textrm{t}}^*(z) dz,
\end{equation}
where $E(z)$ is the scattered field. The target mode position is
kept invariant for simulations with different etch lengths. The emission phase $\phi_\textrm{e} = \angle f_\textrm{e}$ can be used later to adjust the separation between neighbor grating trenches with different etch lengths.

\textit{Step 3: Determine the scattering strength.}

With the mapping from the grating structures to the scattering strength and the parameters defined by the fabrication technology, we can find the optimal scattering strength using the algorithm presented in Sec. \ref{sec:theory}. 

\begin{figure}
    \centering
    \includegraphics[width = 0.64\linewidth]{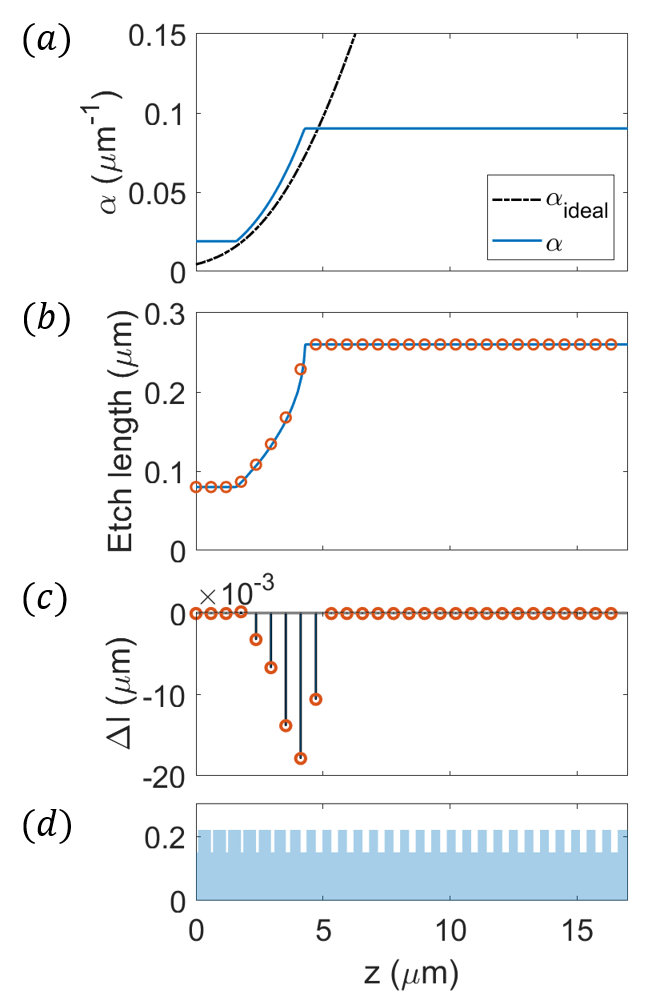}
    \caption{(a) The optimal scattering strength given by our model (solid blue curve), in comparison with the {ideal model} (dashed black curve), which grows monotonically with $z$. (b) The etch length of the grating obtained directly from the scattering strength (solid blue curve) and for each trench in the grating (orange circle). (c) The adjustment of the separation between neighbor grating trenches. (d) Cross section of the grating.}
    \label{fig:design_SOI}
\end{figure}

For the chosen technology, the mapping shown in Fig. \ref{fig:mapping}(c) indicates that $\alpha_\textrm{min} = 0.02$ $\mu$m$^{-1}$, which is limited by the 80 nm minimal feature size, and $\alpha_\textrm{max} = 0.09$ $\mu$m$^{-1}$, which is achieve by etch length $l_e = 0.26$ $\mu$m. The grating length is chosen to be longer than the Gaussian beam size ($L=17$ $\mu$m). Using these parameters, our model gives the optimal scattering strength, which is shown in Fig. \ref{fig:design_SOI}(a), and the optimal beam center position, which is at $z=6.3$ $\mu$m.
As a comparison, we also show the scattering strength given by the {ideal model}, illustrated by the black dashed curve in Fig. \ref{fig:design_SOI}(a). It grows monotonically and exceeds the upper bound of the scattering strength even before the beam center. 

\textit{Step 4: Determine each etch length and position.}

Once the optimal scattering strength is obtained, we use the mapping (Fig. \ref{fig:mapping}(c)) to retrieve the corresponding etch length at $z$. Since the minimal feature size is 80 nm and the maximal scattering strength is achieved at etch length 260 nm, we only use the mapping for $80 \textrm{nm}\leq l_e \leq 260 \textrm{ nm}$. The solid curve in Fig. \ref{fig:design_SOI}(b) shows the etch length at all possible positions along the grating coupler. 

To fix the position of each grating trench, we start from $z=0$, choose the corresponding etch length predicted by our model, and use the estimated grating pitch (\ref{eq:pitch_analytical}) to get the position of the next grating trench, till the end of the grating. Additional care need to be taken when the neighbor grating trenches have different etch lengths, due to their different emission phases. 
When the adjacent trenches, denoted as the $i^{\textrm{th}}$ and $(i+1)^{\textrm{th}}$ trenches, have different emission phases, $\phi_{e,i}$ and $\phi_{e, i+1}$, their separation should have a small deviation from $\Lambda(z_i) - l_e(z_i)$ to restore the phase matching condition. This small change ($\Delta l$) is 
\begin{equation}
\label{eq:delta_l}
    \Delta l_i = \frac{\lambda}{2\pi (n_{\textrm{wg}} - n_\textrm{c}\sin\theta)}(\phi_{e,i} - \phi_{e, i+1}).
\end{equation}
With this correction, the separation between adjacent trenches becomes $\Lambda(z_i) - l_e(z_i) + \Delta l_i$. This adjustment of the grating trench separation is shown in Fig. \ref{fig:design_SOI}(c) for the demonstrated example. After this adjustment, we obtain the position and etch length of each grating trench, as shown in Fig. \ref{fig:design_SOI}(b), which completes the design procedure. A cross section view of the grating is shown in Fig. \ref{fig:design_SOI}(d) for illustration.
Similar grating designs where an apodized section is followed by a uniform grating has been presented in \cite{chen2010apodized}. Our model here provides a rigorous foundation for such designs.

\textit{Post verification and further optimization}

Numerical simulation of this deterministically designed grating coupler shows a coupling efficiency of 61.4\%, which is comparable to grating couple designs with sophisticated optimization \cite{bozzola2015optimising}. The major loss is due to the scattering towards the substrate, which can be suppressed by improving the grating directivity, such as using a thicker silicon core layer, placing a bottom mirror, or using slant gratings or two grating layers \cite{bozzola2015optimising, marchetti2017high, wang2005embeddedslant, ding2014fully, su2018fully, michaels2018inverse, van2007efficient}. To further optimize the grating coupler, one can use this deterministic design as an initial point, since a good initial guess is usually important for many optimization algorithms \cite{michaels2018inverse}.

\section{Two-dimensional apodized non-focusing grating couplers}
\label{sec:2d_nonfocus}

In this section, we investigate the optimal scattering strength in a two-dimensional non-focusing grating coupler, as shown in Fig. \ref{fig:schematic}(a), 
which can be used to couple to high-order modes of a multi-mode fiber or vortex beams.
In these cases, the apodization can be different at different transverse positions, which implies that the grating trenches in Fig. \ref{fig:schematic}(a) may not be parallel or continuous. We discuss the {ideal model} and our model in sequence.

We assume that the power flow of the guided mode is along $z$-direction with no {power flow} in the transverse $x$-direction. This assumption is approximately valid in non-focusing grating couplers. Same as in Secs. \ref{sec:review} and \ref{sec:theory}, we focus only on the amplitude matching and assume that the phase matching and polarization matching are satisfied. Thus, the coupling efficiency is given by (\ref{eq:couplingefficiency2}) and terms in the integrand ($A(x, z)$ and the $A_\textrm{t}(x, z)$) are real. The coupling efficiency is maximized when the overlap integration $\int dz A(x, z) A_\textrm{t}(x, z)$ is maximized at each $x$. Therefore, the problem of finding the optimal scattering strength can be divided into sub-problems of finding the optimal scattering strength of a one-dimensional grating at each $x$-cut.

The scattering strength of the {ideal model} is a direct extension of (\ref{eq:alpha_1}):
\begin{align}
    \label{eq:alpha_2d_1}
    \alpha(x, z) = \frac{1}{2}\frac{S_\textrm{t}(x, z)}{P(x, -\infty) - \int_{-\infty}^z S_\textrm{t}(x, t)dt},
\end{align}
where $P(x, -\infty)$ is the guided power per unit length before the grating region and at transverse position $x$, and the total guided power is $P_\textrm{wg} = \int dx P(x, -\infty)$. Equation (\ref{eq:alpha_2d_1}) should be applied with caution, since it is not obvious that multiplying $S_\textrm{t}(x, z)$ by a constant has no influence on $\alpha(x, z)$. Unlike in the one-dimensional counter-part (\ref{eq:alpha_1}) where $P(-\infty)$ can be re-scaled depending on $S_\textrm{t}(z)$, in the two-dimensional grating, the initial guided power at different transverse positions are related and hence cannot be scaled independently. In fact, to use (\ref{eq:alpha_2d_1}), one should take $S_\textrm{t}(x, z) = b(x)|A_\textrm{t}(x, z)|^2$ where the coefficient $b(x)$ depends on $P(x, -\infty)$. 
Therefore, we proceed to transform (\ref{eq:alpha_2d_1}) into a form such that the scattering strength $\alpha(x, z)$ is invariant when $S_\textrm{t}(x, z)$ is multiplied by a constant, and take $S_\textrm{t}(x, z) = |A_\textrm{t}(x, z)|^2$.

In the case that all the guided power is extracted, $P(x, -\infty) = \int_{-\infty}^{+\infty}S_\textrm{t}(x, t)dt$, the scattering strength is 
\begin{equation}
    \label{eq:alpha_2d_complete}
    \alpha(x, z) = \frac{1}{2}\frac{S_\textrm{t}(x, z)}{\int_z^{+\infty}S_\textrm{t}(x, t)dt}.
\end{equation}
The scattered intensity distribution is 
\begin{align}
    \label{eq:S_2d_nonfocus_complete}
    S(x, z) = P(x, -\infty)\frac{S_\textrm{t}(x, z)}{\int_{-\infty}^{+\infty} S_\textrm{t}(x, t) dt}
\end{align}
We assume the portion of the scattered power that is scattered towards the side of the target mode is $\xi$. Then, the amplitude of the scattered field is $A(x, z) = \sqrt{\xi S(x, z)}$. Substituting into (\ref{eq:couplingefficiency2}), the coupling efficiency with such optimal scattering strength is 
\begin{align}
    \label{eq:coupling_efficiency_2d_complete}
    \eta = \frac{\xi}{P_\textrm{wg}P_\textrm{t}}\Big| \int dx \sqrt{P(x, -\infty)} \sqrt{\int_{-\infty}^{+\infty} dz S_\textrm{t}(x, z)}\Big|^2
\end{align}
Equation (\ref{eq:coupling_efficiency_2d_complete}) can be utilized to find the optimal waveguide cross section in the transverse direction. 

When only part of the guide power is extracted, the extension of (\ref{eq:alpha_partial_scattering}) to the two-dimensional model is non-trivial but has not been considered thoroughly in previous studies. To achieve the amplitude matching between the scattered field and the target mode, the portion of the guided power scattered by the grating ($\zeta$) should be a function of the transverse position:
\begin{align}
    \label{eq:zeta_x}
    \zeta(x) = \frac{\int_{-\infty}^{\infty}S(x, t)dt}{P(x, -\infty)}.
\end{align}
To match $S(x, z)$ with $S_\textrm{t}(x, z)$, the choice of $\zeta(x)$ can be obtained in the following approach. Define $\zeta_\textrm{t}(x)$ as
\begin{align}
    \label{eq:zeta_t}
    \zeta_\textrm{t}(x) = \frac{\int_{-\infty}^{\infty}S_\textrm{t}(x, t)dt}{P(x, -\infty)}. 
\end{align}
We have the freedom of choosing the portion of scattered power at a specific transverse position, for instance, at $x = x_\textrm{m}$, set  $\zeta(x_\textrm{m}) = \zeta_\textrm{m}$. Then, $\zeta(x)$ is fixed through
\begin{align}
    \label{eq:zeta_choice}
    \zeta(x) = \zeta_\textrm{m}\frac{\zeta_\textrm{t}(x)}{\zeta_\textrm{t}(x_\textrm{m})}.
\end{align}
A practical choice of $x_\textrm{m}$ is to choose the maximum of $\zeta_\textrm{t}(x)$ in the range where $P(x, -\infty)$ is also substantial. Near the transverse edge of the grating, the guided power, which appears in the demoninator of (\ref{eq:zeta_t}), approaches zero, and
$\zeta(x)$ determined by (\ref{eq:zeta_choice}) may be larger than unity. Nevertheless, the apodization design is insignificant in those regions due to the negligible scattering power. In those regions, we can set $\zeta(x) \lesssim 1$ if $\zeta(x)$ given by (\ref{eq:zeta_choice}) is larger than one. The portion of the total scattered power is 
\begin{align}
    \label{eq:zeta_total}
    \zeta = \frac{1}{P_\textrm{wg}} \int dx \zeta(x) P(x, -\infty).
\end{align}
With this choice of $\zeta(x)$, the scattering strength is
\begin{align}
    \label{eq:alpha_2d_partial}
    \alpha(x, z) = \frac{1}{2}\frac{\zeta(x)S_\textrm{t}(x, z)}{\int_{-\infty}^{+\infty}S_\textrm{t}(x, t)dt - \zeta(x)\int_{-\infty}^z S_\textrm{t}(x, t)dt},
\end{align}
which is the two-dimensional extension of (\ref{eq:alpha_partial_scattering}). By tuning $\zeta_\textrm{m}$, one can ensure that the scattering strength is below an upper bound. The scattering strength obtained by (\ref{eq:alpha_2d_partial}) ensures amplitude matching and maximal coupling efficiency for the given portion of the scattered power ($\zeta$). The coupling efficiency is $\eta \lesssim \xi \zeta$. It is generally not the optimum when the coupling efficiency is the only objective.

The extension of our model, (\ref{eq:alpha_opt}) and (\ref{eq:alphaconstraint}), to two-dimensional non-focusing gratings is straightforward. Due to the assumption of no power exchange along the transverse $x$-direction, the scattering strength $\alpha(x, z)$ at different $x$ can be obtained independently using (\ref{eq:alpha_opt}) and (\ref{eq:alphaconstraint}). 
We demonstrate such a two-dimensional non-focusing grating coupler design in Sec. \ref{sec:design_oam}.

\section{Two-dimensional grating coupling to a vortex beam}
\label{sec:design_oam}

The two-dimensional extension of our model provides an approach to design apodized grating couplers for sophisticated beams. 
To demonstrate this capability, we study in this section a design of a two-dimensional apodized grating coupler that couples to a Laguerre-Gaussian beam with orbital angular momentum. Such grating couplers are studied previously \cite{zhou2019ultra, nadovich2016forked, nadovich2017focused, liu2016chip}, but the coupling efficiency is still much lower than the coupling efficiency with the simple Gaussian beam. We demonstrate here that the coupling efficiency with a vortex beam is 57.6\%, only a few percent lower than the 61.4\% coupling efficiency with a Gaussian beam shown in Sec. \ref{sec:design_procedure}.

We choose exactly the same technology consideration as discussed in Sec. \ref{sec:design_procedure}. The Laguerre-Gaussian beam radius is $w_0 = 5.2$ $\mu$m and the topological charge of the beam is $l = 1$. The width of the input waveguide is $20$ $\mu$m ($-10 \textrm{ } \mu \textrm{m} \leq x \leq 10 \textrm{ }\mu$m), which matches the target beam in the $x$-direction best. The length of the grating along $z$-direction is also 20 $\mu$m ($0 \leq z \leq 20$ $\mu$m). The beam center is located at $x_0=0, \, z_0=7$ $\mu$m.

The same design procedure as outlined in Sec. \ref{sec:design_procedure} is carried out for the design at each $x$. The only change is in \textit{Step 4} when the separation between neighbor trenches is adjusted to achieve phase matching. For the vortex beam, the target mode has an additional phase
\begin{align}
    \label{eq:phase_oam}
    \psi(x, z) = l\varphi(x, z),
\end{align}
where $l$ is the topological charge and $\varphi(x, z) = \textrm{atan}[(x - x_0)/(z - z_0)]$. The change in separation becomes 
\begin{align}
    \label{eq:delta_l_oam}
    \Delta l_i(x) =  \frac{[\phi_{e,i}(x) - \psi_i(x)] - [\phi_{e, i+1}(x) - \psi_{i+1}(x)]}{2\pi(n_\textrm{wg} - n_\textrm{c}\sin\theta)}\lambda,
\end{align}
where $\psi_i(x) = \psi(x, z_i)$. Equation (\ref{eq:delta_l_oam}) is an extension of (\ref{eq:delta_l}) when the target mode has an extra phase in addition to the phase due to the non-zero incident angle $\theta$. 
The position of the starting grating trench at each \textit{x}-cut ($z_0(x)$) is chosen such that the phase $2\pi(n_\textrm{wg} - n_\textrm{c}\sin\theta)z_0(x)/\lambda - \psi(x, z)$ is independent of $x$, which ensures that the phase matching is satisfied along the transverse direction.

\begin{figure}
    \centering
    \includegraphics[width=0.96\linewidth]{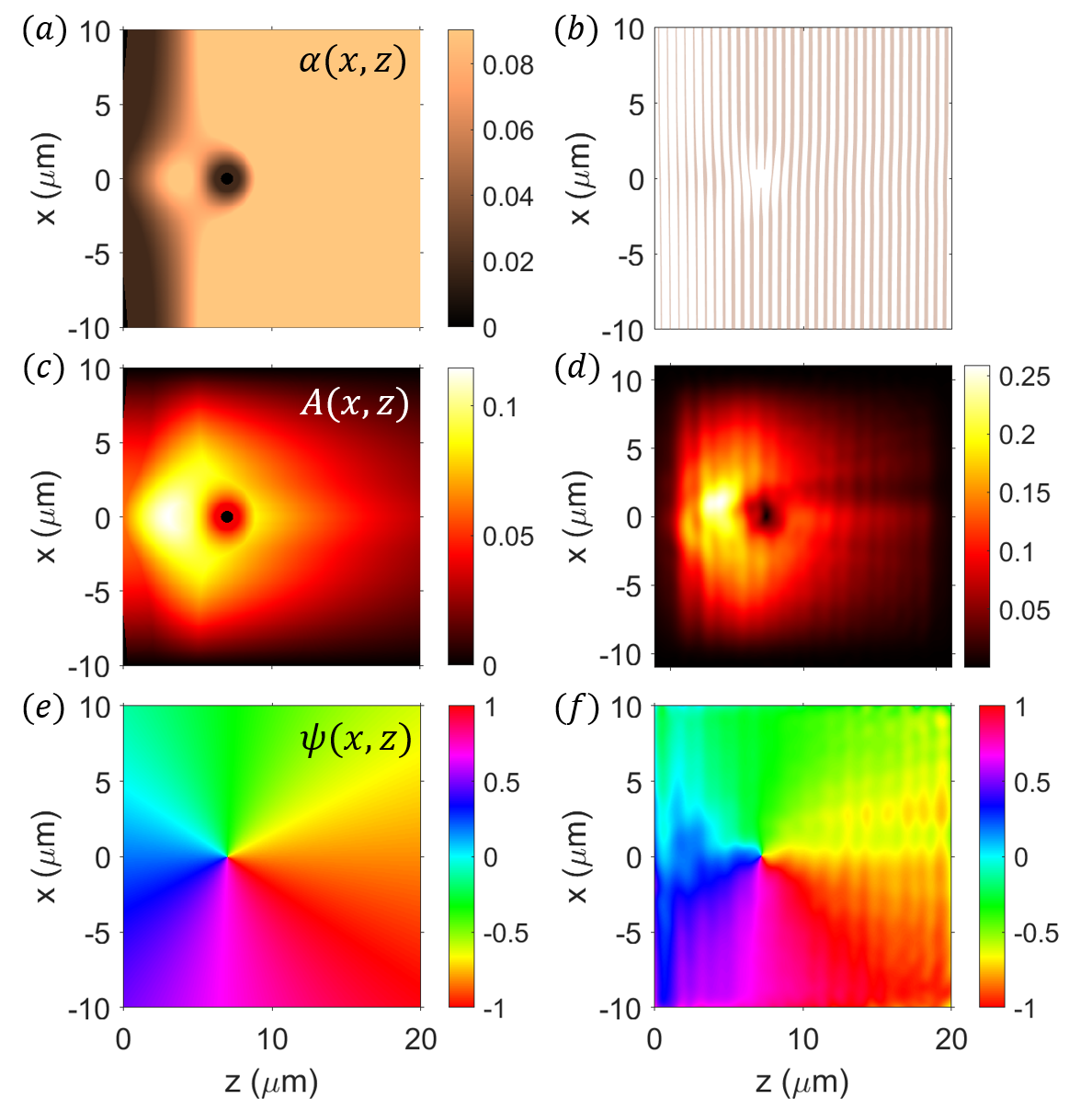}
    \caption{Grating coupler design for Laguerre-Gaussian beam with $l=1$. (a) The optimal scattering strength obtained from our model. (b) The etched trenches on the grating coupler. (c) The amplitude distribution associated with the scattering strength in (a). (d) The amplitude of $E_x$ at a plane 1 $\mu$m above the grating. (e) The additional phase $\psi$ of the Laguerre-Gaussian beam. (f) The phase of $E_x$ at a plane 1 $\mu$m above the grating subtracting $2\pi n_\textrm{c} \sin \theta z/\lambda$.}
    \label{fig:design_oam}
\end{figure}

The scattering strength obtained by our model is shown in Fig. \ref{fig:design_oam}(a). The scattering amplitude $A(x, z) = \sqrt{S(x, z)}$ associated with this scattering strength is presented in Fig. \ref{fig:design_oam}(c). Since our objective is to maximize the coupling efficiency, rather than preserving the mode fidelity, the amplitude distribution is different from the ``donut'' distribution of the Laguerre-Gaussian mode with $l = 1$. 
After determining each etch length from the scattering strength and each trench position at different $x$, we obtain the shape and position of each grating trench, as shown in Fig. \ref{fig:design_oam}(b). 

As a validation of the design procedure, we simulate the designed device using three-dimensional finite-difference time-domain method. The grating coupler is excited by the fundamental guided mode of a 20 $\mu$m wide and 220 nm thick silicon waveguide. We monitor the field at 1 $\mu$m above the grating. The amplitude and phase of the transverse electric field $E_x$ are respectively shown in Fig. \ref{fig:design_oam}(d) and (f), where the phase associated with the nonzero incident angle is subtracted. The phase of the scattered field matches well with the phase of the target beam, shown in Fig. \ref{fig:design_oam}(e). In the simulation, we observe that 62.4\% power is scattered upwards and the coupling efficiency is $\eta = 57.6$\%. The mode matching efficiency, which is the ratio between coupling efficiency and the portion of light scattered upwards, is as high as 92\%. Comparing the coupling efficiency with the one-dimensional grating coupler coupling with a simple Gaussian beam ($\eta = 61.4$\%), the reduction in coupling efficiency is only 3.8 percentage point. This coupling efficiency is also significantly higher than previous design and demonstration of coupling efficiency $\sim$ 32\% for the vortex beam \cite{nadovich2017focused, nadovich2016forked}. This demonstration suggests that our model and design procedure indeed generate highly efficient grating couplers, and can be applicable for achieving efficient coupling to complicated target modes.

\section{Two-dimensional apodized focusing grating couplers}
\label{sec:2d_focusing}

\begin{figure}
    \centering
    \includegraphics[width = 0.6\linewidth]{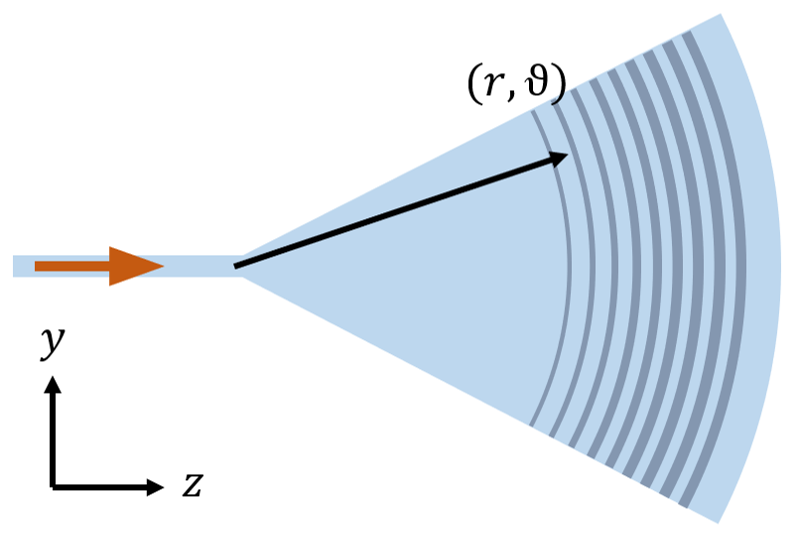}
    \caption{Schematic of a focusing apodized grating coupler with a top-down view. The darker blue part indicates the etched grating trenches. The input waveguide connects with the fan-shape slab at the coordinate origin.  $r$ represents the distance to the origin, and $\vartheta$ is the angle measured from $z$-axis.}
    \label{fig:schematic_fanout}
\end{figure}

In this section, we consider the design of coupler that couples from a single-mode on-chip waveguide to a free-space beam. In silicon photonics, due to the high index of silicon, such a single-mode waveguide typically has a corss-section of approximately 0.5 $\mu$m, which is far smaller than the size of the free-space beam. In principle, one can first use a waveguide taper structure to expand the waveguide mode on-chip, and then use the non-focusing grating coupler that we have already designed to couple to a free-space beam. However, such a taper structure typically occupies substantial chip area \cite{mekis2010grating}. Here, instead, we consider a more compact design where a focusing coupler is used to directly couple from a single-mode on-chip waveguide to a free-space beam.


In a typical focusing grating coupler, the single-mode waveguide is connected to a fan-shape free-propagation region and the grating trenches are located on curves close to concentric ellipses rather than parallel lines as in non-focusing grating couplers \cite{van2007compact, mehta2017precise, nadovich2017focused}. A schematic of such a focusing grating coupler is illustrated in Fig. \ref{fig:schematic_fanout}. In this section, we extend both the {ideal model} and our model to find the optimal scattering strength for the focusing grating. We believe that such extension is useful practically, but it has not been considered systematically in previous studies \cite{mehta2017precise, nadovich2017focused}.

The power flow in the fan-shape region is almost along the radial direction, so we assume that the power exchange between different angles $\vartheta$ is zero.  With this assumption, we can treat the apodization design for different angles separately.
Same as in Sec. \ref{sec:2d_nonfocus}, the phase matching and polarization matching are omitted and we investigate only the amplitude matching, since the phase matching in focusing grating couplers is well-studied \cite{mekis2010grating, becker2019out, marchetti2019coupling}. The coupling efficiency is 
\begin{align}
    \label{eq:couplingefficiency_focusing}
    \eta = \frac{1}{P_\textrm{wg}P_\textrm{t}}\Big| \int d\vartheta \int r A(r, \vartheta) A_\textrm{t}(r, \vartheta) dr \Big|^2,
\end{align}
where the amplitudes $A(r, \vartheta) = A(x, z)$ and the coordinates $z = r\cos\vartheta, x = r\sin\vartheta$. 

The total guided power on a constant radius surface in the fan-shape region is 
\begin{equation}
    \label{eq:power_r}
    P(r) = \int r P(r, \vartheta) d\vartheta,
\end{equation}
where $P(r)$ is the total guide power at $r$, and $P(r, \vartheta)$ is the power per unit length at position $(r, \vartheta)$. Suppose that the grating region starts at $r = r_0$. The power decay along the radial direction is 
\begin{align}
    \label{eq:power_decay_r}
    \frac{\partial}{\partial r}[r P(r, \vartheta)] = -2\alpha(r, \vartheta) r P(r, \vartheta).
\end{align}
The scattered power per unit area is 
\begin{equation}
    \label{eq:scatter_r_theta}
    S(r, \vartheta) = 2\alpha(r, \vartheta)P(r, \vartheta).
\end{equation}
To maximize the coupling efficiency, the scattered power should match the target mode, i.e. $S(r, \vartheta) = S_\textrm{t}(r, \vartheta)$, where $S_\textrm{t}(r, \vartheta)$ is the target intensity distribution expressed with polar coordinates. With (\ref{eq:power_decay_r}) and (\ref{eq:scatter_r_theta}), we have
\begin{align}
    \label{eq:power_decay_r_target}
    & \frac{\partial}{\partial r}[rP(r,\vartheta)] = -rS_\textrm{t}(r, \vartheta), \\
    \label{eq:power_r_target}
    & P(r, \vartheta) = \frac{r_0 P(r_0, \vartheta)}{r} - \frac{1}{r}\int_{r_0}^r tS_\textrm{t}(t, \vartheta) dt, \\
    \label{eq:alpha_2d_focusing_general}
    & \alpha(r, \vartheta) = \frac{1}{2}\frac{rS_\textrm{t}(r, \vartheta)}{r_0 P(r_0, \vartheta) - \int_{r_0}^r tS_\textrm{t}(t, \vartheta) dt}.
\end{align}
Equation (\ref{eq:alpha_2d_focusing_general}) is an extension of (\ref{eq:alpha_1}) for the one-dimensional grating and (\ref{eq:alpha_2d_1}) for the non-focusing grating. 

Similar to the discussion in Sec. \ref{sec:2d_nonfocus}, we would like to transform (\ref{eq:alpha_2d_focusing_general}) into a scale-invariant form, such that it is transparent that multiplying $S_\textrm{t}(r, \vartheta)$ by a constant does not change $\alpha(r, \vartheta)$. Then we can simply take $S_\textrm{t}(r, \vartheta) = |A_\textrm{t}(r, \vartheta)|^2$ to obtain the scattering strength. 

In the case that all the guided power is scattered, this boundary condition suggests that $r_0 P(r_0, \vartheta) = \int_{r_0}^{+\infty}tS_\textrm{t}(t, \vartheta)dt$, and the scattering strength is 
\begin{align}
    \label{eq:alpha_2d_focusing_complete}
    \alpha(r, \vartheta) = \frac{1}{2}\frac{r S_\textrm{t}(r, \vartheta)}{\int_r^{+\infty}t S_\textrm{t}(r, \vartheta) dt}.
\end{align}
The corresponding scattering intensity distribution is 
\begin{align}
    \label{eq:S_design_2d_focusing}
    S(r, \vartheta) = r_0 P(r_0, \vartheta)\frac{S_\textrm{t}(r, \vartheta)}{\int_{r_0}^{+\infty} t S_\textrm{t}(t, \vartheta) dt}.
\end{align}
Recalling that $A(r, \vartheta) = \sqrt{\xi S(r, \vartheta)}$ and $A_\textrm{t}(r, \vartheta) = \sqrt{S_\textrm{r}(r, \vartheta)}$, the coupling efficiency (\ref{eq:couplingefficiency_focusing}) is:
\begin{align}
    \label{eq:coupling_efficiency_complete_focusing}
    \eta = \frac{\xi}{P_\textrm{wg}P_\textrm{t}}\Big | \int d\vartheta \sqrt{r_0P(r_0, \vartheta)} \sqrt{\int_{r_0}^{+\infty} r S_\textrm{t}(r, \vartheta)dr}\Big|^2.
\end{align}
The open angle of the fan-shape region and the position of the target beam center are important design parameters for the focusing grating couplers. The optimal open angle with respect to the beam center position can be found using (\ref{eq:coupling_efficiency_complete_focusing}), which is demonstrated in Sec. \ref{sec:implication}.

When only part of the guided power is scattered out and the scattering strength is chosen to maximize the mode matching, the portion of scattered power should be a function of the angle. The treatment is similar to that in Sec. \ref{sec:2d_nonfocus}. Let 
\begin{align}
    \label{eq:zeta_2d_target}
    \zeta_\textrm{t}(\vartheta) = \frac{\int_{r_0}^{+\infty} tS_\textrm{t}(t, \vartheta)dt}{r_0 P(r_0, \vartheta)}.
\end{align}
We have the freedom to choose the portion of scattered power at a specific angle, for instance $\zeta(\vartheta_\textrm{m}) = \zeta_\textrm{m}$. Then, the scattered portion as a function of angle is \begin{align}
    \label{eq:zeta_2d_design}
    \zeta(\vartheta) = \zeta_\textrm{m}\frac{\zeta_\textrm{t}(\vartheta)}{\zeta_\textrm{t}(\vartheta_\textrm{m})}.
\end{align}
The practical choice of $\vartheta_\textrm{m}$ and the subtleties of adjusting $\zeta(\vartheta)$ near the edge of the fan-shape region are similar to the discussion in Sec. \ref{sec:2d_nonfocus}.
The portion of the total scattered power is 
\begin{equation}
    \label{eq:zeta_total_2d_focusing}
    \zeta = \frac{1}{P_\textrm{wg}}\int d\vartheta \zeta(\vartheta)r_0 P(r_0, \vartheta).
\end{equation}
Using the boundary condition $\zeta(\vartheta) r_0 P(r_0, \vartheta) = \int_{r_0}^{+\infty}t S_\textrm{t}(t, \vartheta)dt$, the scattering strength is 
\begin{align}
    \label{eq:alpha_2d_focusing_partial}
    \alpha(r, \vartheta) = \frac{1}{2}\frac{\zeta(\vartheta) r S_\textrm{t}(r, \vartheta)}{\int_{r_0}^{+\infty} t S_\textrm{t}(t, \vartheta) dt - \zeta(\vartheta) \int_{r_0}^{r}t S_\textrm{t}(t, \vartheta) dt}.
\end{align}
The corresponding coupling efficiency is $\eta \lesssim \xi \zeta$.

Our model can also be extended to the design of apodized focusing gratings. From (\ref{eq:power_decay_r}) and (\ref{eq:scatter_r_theta}), the scattering intensity associated with a scattering strength distribution is
\begin{align}
    \label{eq:scattering_2d_focusing}
    S(r, \vartheta) = 2\alpha(r, \vartheta) \frac{r_0}{r}P(r_0, \vartheta)\exp\Big[-2\int_{r_0}^{r} \alpha(t, \vartheta) dt\Big].
\end{align}
The coupling efficiency given by (\ref{eq:couplingefficiency_focusing}) is
\begin{align}
    \label{eq:coupling_efficiency_2d_ourmodel}
    \begin{split}
        \eta = &\, \frac{\xi}{P_\textrm{wg}P_\textrm{t}}\Big | \int d\vartheta \sqrt{r_0 P(r_0, \vartheta)} \int_{r_0}^{r_0+L}\sqrt{2\alpha(r, \vartheta)}\\
        &\cdot \exp\Big[ -\int_{r_0}^r \alpha(t, \vartheta)dt \Big] \sqrt{r} A_\textrm{t}(r, \vartheta)dr \Big|^2,
    \end{split}
\end{align}
where we assume that the grating lies between $r_0$ and $r_0+L$, and $A(r, \vartheta) = \sqrt{\xi S(r, \vartheta)}$ is used. Since each term in the absolute sign in (\ref{eq:coupling_efficiency_2d_ourmodel}) is real, the maximal coupling efficiency is achieved when the following term is maximized for each $\vartheta$.
\begin{align}
    \label{eq:f_2d_focusing}
    \begin{split}
    f(\vartheta) = & \, \int_{r_0}^{r_0 + L} \sqrt{2\alpha(r, \vartheta)}\exp \Big[-\int_{r_0}^r \alpha(t, \vartheta) dt \Big] \\ & \cdot \sqrt{r} A_\textrm{t}(r, \vartheta) dr.
    \end{split}
\end{align}
The physical intuition is that the scattering strength along each angle can be designed independent of other angles. 
Comparing to (\ref{eq:f}), the only difference in (\ref{eq:f_2d_focusing}) is to replace $A_\textrm{t}(z)$ by $\sqrt{r}A_\textrm{t}(r, \vartheta)$, besides trivial shift of origin and change of integration variable. These changes does not influence the optimal substructure of the optimization problem outlined in (\ref{eq:optimization}). Therefore, the optimal scattering strength can be found following the derivation in Sec. \ref{sec:theory}. The optimal scattering strength is 
\begin{align}
    \label{eq:alpha_opt_2d_focusing}
    \alpha_r(\vartheta) = \frac{r A_\textrm{t}^2(r, \vartheta)}{2\Big\{ \int_r^{r_0+L}e^{-\int_r^s\alpha^\star(t)dt}\sqrt{2\alpha^\star(s, \vartheta)s}  A_\textrm{t}(s, \vartheta)ds \Big\}^2}, 
\end{align}
with the same adjustments outlined in (\ref{eq:alphaconstraint}) to explicitly take the upper and lower bounds of the scattering strength into account.

\section{Implications of our model}
\label{sec:implication}
Equipped with our model, we can study how the upper limit of the coupling efficiency depends on the upper and lower bounds of the scattering strength and the grating length. With the extension of the {ideal model} to the two-dimensional focusing gratings, we also investigate the optimal open angle of the fan-shape region with respect to the beam center position.
For practical purposes, we choose the target mode in this section to be a Gaussian beam with beam radius $w_0 = 5.2$ $\mu$m, which resembles the mode in a single-mode fiber.

\begin{figure}[h]
    \centering
    \includegraphics[width = 0.6\linewidth]{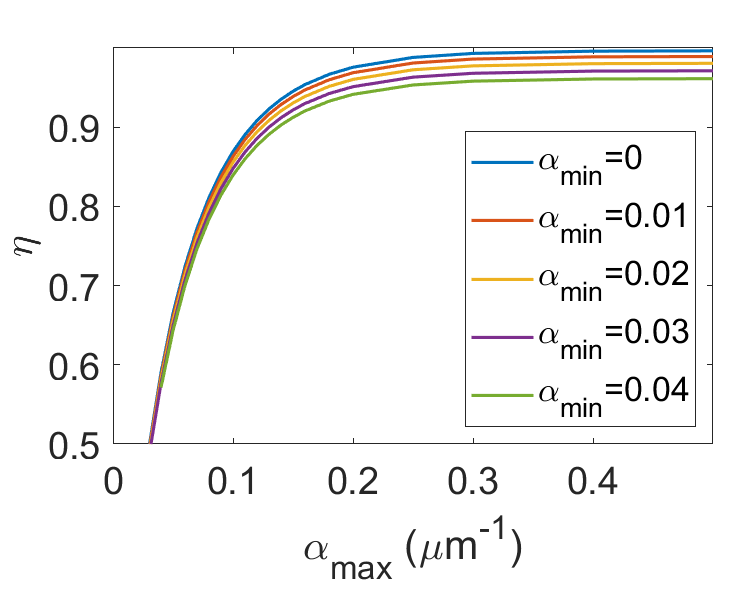}
    \caption{The upper bound of the coupling efficiency limited by the maximal scattering strength, where the length of the grating coupler is much longer than the Gaussian beam waist. Curves with different colors represent different lower bounds of the scattering strength.}
    \label{fig:alpha_sweep}
\end{figure}

We first study how the finite range of scattering strength influences the upper limit of the coupling efficiency. We assume that all the scattered power is towards the target mode side ($\xi=1$) and the mode matching in transverse dimension is perfect. To avoid the influence of finite coupling length, the grating coupler is much longer than the Gaussian beam waist. Figure \ref{fig:alpha_sweep} shows the upper limit of the coupling efficiency as a function of the upper bound of the scattering strength, where different curves represent different lower bounds of the scattering strength.  
The coupling efficiency, including the asymptotic limit ($\alpha_\textrm{max} \rightarrow \infty$), decreases mildly with the lower bound of the scattering strength.
We also find that the coupling efficiency grows with $\alpha_\textrm{max}$ but starts to saturate when $\alpha_\textrm{max}$ is above about 0.15 $\mu\textrm{m}^{-1}$. 
This implies the intrinsic challenge for low-index contrast gratings coupling with single-mode fiber, where $\alpha_\textrm{max}$ is small. Nevertheless, this challenge can be bypassed with techniques to expand the target mode size \cite{chen2016high}. A discussion of how the optimal scattering strength varies with a scaling of the target mode size is presented in Appendix \ref{app:scaling}, which is useful to generalized the above results to different Gaussian beam radius.

\begin{figure}
    \centering
    \includegraphics[width = 0.64\linewidth]{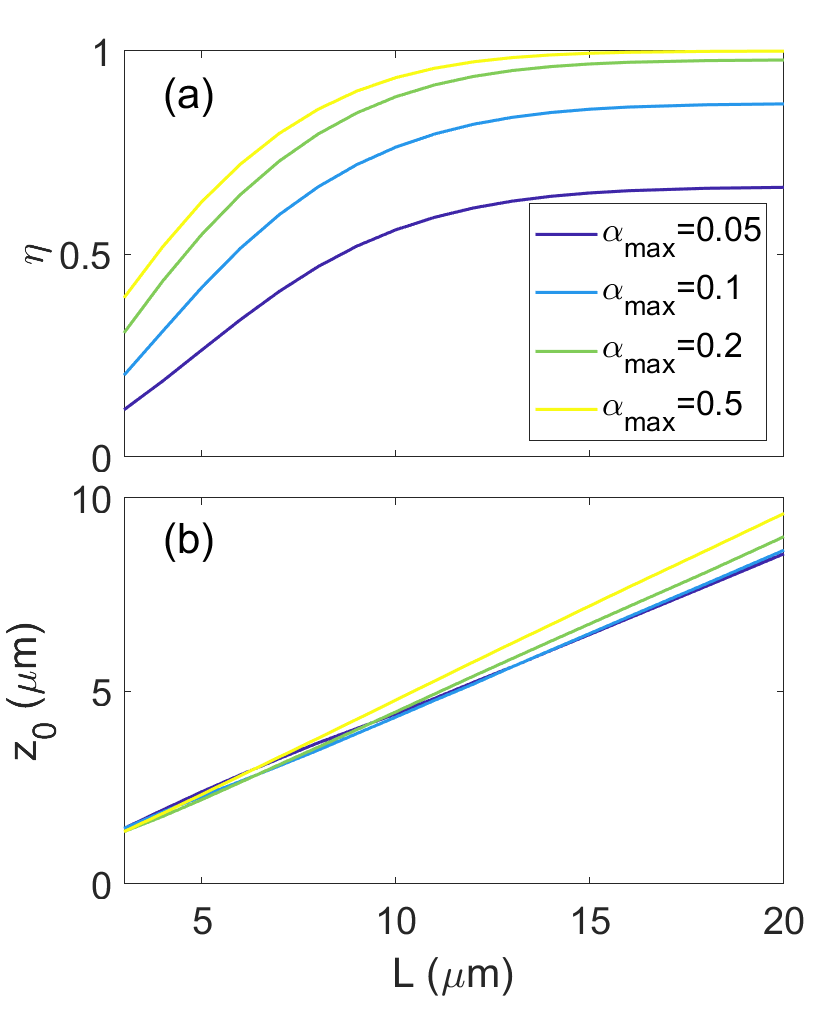}
    \caption{(a) The upper bound of the coupling efficiency limited by the the length of the grating coupler. Curves with different colors represent different upper bounds of the scattering strength. The lower bound of the scattering strength is set to be zero. (b) The corresponding optimal position of the Gaussian beam center.}
    \label{fig:L_sweep}
\end{figure}

To achieve compact grating couplers \cite{zhou2019ultra, sun2013large}, shorter grating length is preferred. Our model also predicts the influence of the finite grating length. Figure \ref{fig:L_sweep}(a) shows the upper limit of the coupling efficiency as a function of the grating length, where different curves represent different $\alpha_\textrm{max}$ while $\alpha_\textrm{min}$ is set to zero. The corresponding optimal position of the Gaussian beam center is shown in Fig. \ref{fig:L_sweep}(b). We find that the coupling efficiency increases with the grating length, and the increment becomes insignificant when the length exceeds about 12 $\mu$m ($L \sim 2.4w_0$). This trend consistently holds for different $\alpha_\textrm{max}$. Therefore, the length of the grating coupler should be larger than $\sim 2.4 w_0$ to avoid significant efficiency decrease. We also observe from Fig. \ref{fig:L_sweep}(a) that large $\alpha_\textrm{max}$ has greater advantage when the grating is shorter. This is consistent with previous studies using full etch and short apodization region for compact grating couplers \cite{sun2013large}.


\begin{figure}
    \centering
    \includegraphics[width=0.7\linewidth]{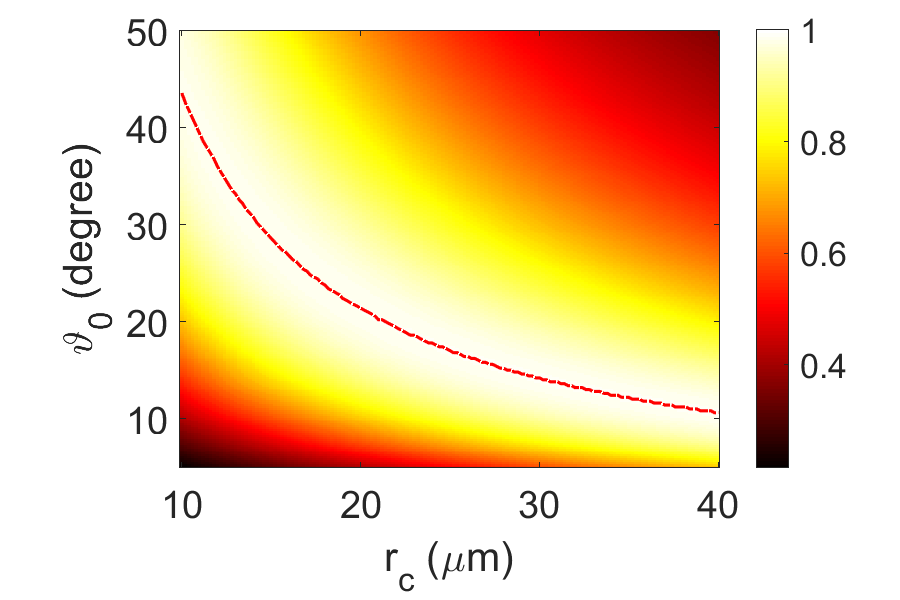}
    \caption{The upper limit of the coupling efficiency given by the {ideal model} for the focusing grating as a function of the position of the Gaussian beam center ($r_c$) and the open angle of the fan-shape region ($\vartheta_0$).}
    \label{fig:fan_angle}
\end{figure}

As a demonstration of the extension of the {ideal model} for the two-dimensional focusing grating, we investigate the optimal open angle as a function of the position of the Gaussian beam center. Suppose that the Gaussian beam center is located at $(r = r_c, \vartheta=0)$, and the open angle of the fan-shape region is $\vartheta_0$, i.e. $\vartheta \in [-\vartheta_0, \vartheta_0]$. The guided power in the fan-shape free propagation region can be approximated by \cite{mehta2017precise}
\begin{align}
    \label{eq:P_wg_focusing}
    P(r_0, \vartheta) = \frac{P_\textrm{wg}}{r_0 \vartheta_0}\cos^2\Big( \frac{\pi \vartheta}{2 \vartheta_0}\Big), \textrm{ } \vartheta \in [-\vartheta_0, \vartheta_0].
\end{align}
The resulting upper limit of the coupling efficiency given by (\ref{eq:coupling_efficiency_complete_focusing}), where $\xi=1$ and the target mode is a Gaussian beam, is shown in Fig. \ref{fig:fan_angle} for different open angles and beam positions. The open angle that maximizes the coupling efficiency for each beam position is highlighted by the red dashed curve. The upper limit of the coupling efficiency is between 98\% and 99\% along this curve. Figure \ref{fig:fan_angle} suggests that the optimal open angle should decrease when the beam center is farther away from the waveguide terminal. This result can function as a guidance in designing focusing grating couplers.

\section{Conclusion}
\label{sec:conclusion}

In conclusion, we explicitly consider the upper and lower bound of the scattering strength in an analytical model for designing apodized grating couplers. We prove that the obtained scattering strength is the global optimum and provide an algorithm to compute the scattering strength numerically. Equipped with our model, we present a deterministic design procedure to generate a highly efficient apodized grating coupler. The demonstrated design in a silicon-on-insulator platform has coupling efficiency comparable with previous designs after sophisticated optimization. We further extend both the previous {ideal model} and our model to two-dimensional non-focusing and focusing gratings. Using the two-dimensional extension of our model, we demonstrate a grating coupler coupling to a vortex beam with topological charge $l=1$. The mode matching efficiency is 92\% and the total coupling efficiency 57.6\% is only 3.8 percentage point lower than the design for a standard Gaussian beam. We finally discuss how the finite scattering strength and coupling length influence the upper bounds of the coupling efficiency of a grating coupler coupling with a single mode fiber. 

Our systematical modeling of the apodized grating coupler can provide guidance to design highly efficient grating couplers in different fabrication technologies and for various target modes. The deterministic design procedure presented in this study can generate highly efficient grating couplers quickly or function as a near-optimum starting point for more advanced optimization.

\appendices
\section{Optimal substructure of the constraint optimization problem}
\label{app:optimal_substructure}
In this Appendix, we prove that the constraint optimization problem described by (\ref{eq:optimization}) exhibits optimal substructure by showing that the optimal scattering strength $\alpha^\star(z)$, which maximizes $f$, must also maximize $f^+(z)$ for any $z \in [0, L]$. 
We prove it by counter-evidence.

Suppose the scattering strength $\alpha^\star(z)$ maximizes $f$ but does not maximize $f^+(z)$ for some $z \in [0, L]$, for instance $z = z_1$. Since $f^+(0) = f$ and $f^+(L)=0$, $z_1$ is in the range $z_1 \in (0, L)$. Based on the presumption, there exists another scattering strength distribution $\alpha_1(s)$ for $s \in [z_1, L]$, such that $f^+(z_1)$ takes a larger value, i.e. $f^+(z_1; a_1) > f^+(z_1; a^\star)$. Hence, we can construct another scattering strength distribution $\alpha_2$ by taking $\alpha^\star$ in $[0, z_1)$ and $\alpha_1$ in $[z_1, L]$, i.e.
\begin{align}
    \label{eq:alpha_construct}
    \alpha_2(z) = 
    \begin{cases}
    \alpha^\star(z) & 0 \leq z < z_1 \\
    \alpha_1(z) & z_1 \leq z \leq L
    \end{cases}.
\end{align}
We then compare the overlap integral $f$ of the scattering strength $\alpha^\star(z)$ and $\alpha_2(z)$. Recall the separation of $f$ into two parts with $z = z_1$:
\begin{align}
    \label{eq:f_separation_repeat}
    f = f^-(z_1) + \exp \Big [{-\int_0^{z_1}\alpha(t)dt} \Big ] f^+(z_1),
\end{align}
where $f^-$ and $f^+$ are given in (\ref{eq:f_minus}) and (\ref{eq:f_plus}) with $z = z_1$. We find that $f^-(z_1)$ and the factor $\exp[-\int_0^{z_1} \alpha(t) dt]$ are the same for the scattering strength $\alpha^\star(z)$ and $\alpha_2(z)$, since the scattering strength in $[0, z_1)$ are the same. Because the scattering strength $\alpha_2(z)$ gives larger $f^+(z_1)$, the overall overlap integral $f$ is also larger, i.e. $f(\alpha_2) > f(\alpha^\star)$. This contradicts with the presumption that the scattering strength $\alpha^\star(z)$ maximizes $f$. Thus, we prove that the scattering strength $\alpha^\star(z)$ that maximizes $f$ must also maximize $f^+(z)$ for any $z \in [0, L]$. The optimization problem (\ref{eq:optimization}) therefore exhibits optimal substructure.

\section{Optimal scattering strength with a scaling of the target mode size}
\label{app:scaling}
In this Appendix, we study how the optimal scattering strength changes when the target mode is transformed by a scaling operation. For simplicity, we only consider the one-dimensional system.
Suppose the target mode after the transformation ($\tilde{S}_\textrm{t}$) is related to the original target mode ($S_\textrm{t}$) by:
\begin{equation}
    \label{eq:target_mode_scaling}
    \tilde{S}_\textrm{t}(z) = S_\textrm{t}(\gamma z),
\end{equation}
where $\gamma$ is the scaling factor. We find that the optimal scattering strength matching the target mode after the scaling transformation ($\tilde{\alpha}(z)$) is related to the original optimal scattering strength ($\alpha(z)$) by:
\begin{align}
    \label{eq:alpha_scaling}
    \tilde{\alpha}(z) = \gamma \alpha (\gamma z).
\end{align}
One can check that this relation holds for the {ideal model} (\ref{eq:alpha_complete_scattering}) and (\ref{eq:alpha_partial_scattering}). It also holds for our model (\ref{eq:alpha_opt}), if the upper and lower bounds of the scattering strength and the length of the grating coupler are transformed accordingly:
\begin{align}
    \label{eq:alpha_min_scale}
    \tilde{\alpha}_\textrm{min} & = \gamma \alpha_\textrm{min}, \\
    \tilde{\alpha}_\textrm{max} & = \gamma \alpha_\textrm{max}, \\
    \tilde{L} & = L/\gamma,
\end{align}
where the quantities with a tilde are those matching the target mode after the scaling transformation.



\section*{Acknowledgment}
The authors would like to thank Dr. Min Teng from imec USA Nanoelectronics Design Center Inc., Mr. Nathan Z. Zhao and Prof. Meir Orenstein for helpful discussions. We acknowledge support from the ACHIP program, funded by the  Gordon  and  Betty  Moore Foundation  (GBMF4744). Z. Zhao also acknowledges support from Stanford Graduate Fellowship.

\ifCLASSOPTIONcaptionsoff
  \newpage
\fi




\bibliographystyle{IEEEtran}
\bibliography{grating_coupler_ref.bib}
\end{document}